\documentclass[11pt]{article}
\usepackage{authblk}
\usepackage{amssymb}
\usepackage{amsmath}
\usepackage{amsthm}
\usepackage{wasysym}
\usepackage{lmodern}
\usepackage{graphicx,color,epsfig}
\usepackage{subfigure}

\begin{document}

\title{$b\to ss{\bar d} $ decay in Randall-Sundrum models}
\author[]{Cai-Dian L\"u$\footnote{lucd@ihep.ac.cn}$}
\author[]{Faisal Munir$\footnote{faisalmunir@ihep.ac.cn}$}
\author[]{Qin Qin$\footnote{qin@physik.uni-siegen.de}$}
\affil[]{Institute of High Energy Physics, Chinese Academy of
Sciences, Beijing 100049, China} \affil[]{University of Chinese
Academy of Sciences, Beijing 100049, China}
\date{\today}
\maketitle
\begin{abstract}
The extremely small branching ratio of $b\to ss{\bar d}$ decay in
the Standard Model makes it a suitable channel to explore
new-physics signals. We study this $\Delta S=2$ process in
Randall-Sundrum models, including the custodially protected and the
bulk-Higgs Randall-Sundrum models. Exploring the experimentally
favored parameter spaces of these models, it suggests a possible
enhancement of the decay rate, compared to the Standard Model
result, by at most two orders of magnitude.
\end{abstract}

\section{Introduction}
\label{A1} In studying flavor-changing neutral-current (FCNC)
transitions in rare B decays for exploring  new physics (NP), one
major difficulty is, how to reliably subtract the Standard Model
(SM) background. Theoretical uncertainties in FCNC transitions make
it hard to conclude about definite new physics signals against SM
predictions. For this reason, an alternative approach suggested in
\cite{PhysRevLett.81.4313,Huitu:1998pa} is to consider processes
which have tiny strengths in SM so that mere detection of such
processes will indicate NP. One such process is the rare $b\to
ss{\bar d}$ decay, as reported in
\cite{PhysRevLett.81.4313,Huitu:1998pa}, which can serve the purpose
of exposing NP.

The $\Delta S=2$ $b\to ss{\bar d}$ process is box mediated in SM and
is found to occur with a branching ratio of the order of $10^{-12}$.
The authors of Ref. \cite{PhysRevLett.81.4313} suggested $B^-\to
K^-K^-\pi^+$ as the most appropriate mode for experimental searches
and many other studies of the $b\to ss{\bar d}$ decay have been
conducted in various beyond SM scenarios
\cite{Wu:2003kp,Fajfer:2001ht,Pirjol:2009vz}. The first search was
reported in \cite{Abbiendi:1999st} and
 upper limits were given by both B factories
\cite{Garmash:2003er,Aubert:2003xz,Aubert:2008rr}, with the current
upper limit reported by BABAR Collaboration to be $\mathcal{B}(B^-\to
K^-K^-\pi^+)<1.6\times 10^{-7}$. Moreover, two-body exclusive decays
of $B^-$ \cite{Fajfer:2000ny} and $B_c$ \cite{Fajfer:2004fx}, which
are driven by the $b\to ss{\bar d}$ transition, have also been studied in
SM and in various extensions of it.

In this paper, we consider the inclusive $b\to ss{\bar d}$ decay in
Randall-Sundrum (RS) models \cite{Randall:1999ee,Grossman:1999ra}.
We shall study two models known as the RS model with custodial
protection $(\text {RS}_c)$
\cite{Agashe:2006at,Carena:2006bn,Albrecht:2009xr,Biancofiore:2014wpa,Biancofiore:2014uba}
and the bulk-Higgs RS model \cite{Qin:2016jca,Archer:2014jca}, in
both of which FCNC transitions occur at tree level.

\section{$\text{RS}$ model with custodial protection}
\label{A2} $\text{RS}_c$ model is based on a single warped
extra-dimension with the bulk gauge group
$\text{SU}(3)_c\times\text{SU}(2)_L\times\text{SU}(2)_R\times\text{U}(1)_X\times
P_{LR}$. In the $\text{RS}_c$ model, the $\Delta S=2$ $b\to ss{\bar
d}$ decay receives tree level contributions from the Kaluza-Klein
(KK) gluons, the heavy KK photons, new heavy electroweak (EW) gauge
bosons $Z_H$ and $Z^{\prime}$, and in principle the $Z^0$ boson.
Custodial protection of the $Zb_L{\bar b}_L$ coupling through the
discrete $P_{LR}$ symmetry in order to satisfy EW precision
constraints render tree-level $Z^0$ contributions to be negligible.
It was pointed out in \cite{Bauer:2009cf} that for the $\text{RS}_c$
model the $\Delta F=2$ contributions from Higgs boson exchange are
of $\mathcal O(v^4/{M_{\text{KK}}}^4)$ ($v\approx$ 246 GeV is the
Higgs vacuum expectation value and $M_\mathrm{KK}$ is the KK scale
always larger than 1 TeV) and the importance of Higgs FCNCs is
limited with the most pronounced effects occurring in the case of
the CP-violating parameter $\epsilon_K$, but even there they are
typically smaller than the corrections due to KK-gluon exchanges
\cite{Duling:2009pj}. Therefore, in view of the possible Higgs-boson
effects to be insignificant in $\Delta F=2$ processes, we simply
neglect them in our study of the $b\to ss{\bar d}$ decay in the
$\text{RS}_c$ model.

For the $\text{RS}_c$ model, we consider only first KK excitations
of gauge bosons with $M_{\text {KK}}$ setting the mass scale for the
low-lying KK excitations of the SM particles such that the mass of
the first KK bosons are given by $M_{g^{(1)}}\approx2.45$ $M_{\text
{KK}}$. Here it is important to mention that we have used a
different notation for the mass of the first KK gluon than in
\cite{Blanke:2008zb}, our $M_{\text {KK}}$ corresponds to their $f$.
The dominant contribution comes from the KK gluon, while the new
heavy EW gauge bosons $(Z_H, Z^{\prime})$ can compete with it. The
tree-level $Z^0$ and KK photon contributions are very small. The
effective Hamiltonian for the $\Delta S=2$ $b\to ss{\bar d}$ decay
mediated by exchanges of the lightest KK gluon, the lightest KK
photon and  $(Z_H, Z^{\prime})$ with the Wilson coefficients
corresponding to $\mu=\mathcal{O}(M_{g^{(1)}})$ is given by
\begin{align}
\label{01}
[\mathcal{H}_{\text{eff}}^{\Delta S=2}]_{\text{KK}}&=\frac{1}{2(M_{g^{(1)}})^2}[C_1^{VLL}\mathcal{Q}_1^{VLL}+C_1^{VRR}\mathcal{Q}_1^{VRR}\notag\\
&+C_1^{LR}\mathcal{Q}_1^{LR}+C_2^{LR}\mathcal{Q}_2^{LR}+C_1^{RL}\mathcal{Q}_1^{RL}+C_2^{RL}\mathcal{Q}_2^{RL}],
\end{align}
where
\begin{align}\label{02}
\mathcal{Q}_1^{VLL}&=(\bar s\gamma_{\mu}P_L b)(\bar s\gamma^{\mu}P_L
d),\notag\\
\mathcal{Q}_1^{VRR}&=(\bar s\gamma_{\mu}P_R  b)(\bar
s\gamma^{\mu}P_R
d),\notag\\
\mathcal{Q}_1^{LR}&=(\bar s\gamma_{\mu}P_L b)(\bar
s\gamma^{\mu}P_Rd),\notag\\
\mathcal{Q}_2^{LR}&=(\bar s P_L b)(\bar s P_R d),\notag\\
\mathcal{Q}_1^{RL}&=(\bar s\gamma_{\mu}P_R b)(\bar
s\gamma^{\mu}P_Ld),\notag\\
\mathcal{Q}_2^{RL}&=(\bar s P_R b)(\bar s P_L d),
\end{align}
and
\begin{align}\label{03}
C_i^{j}(M_{g^{(1)}})&=[C_i^{j}(M_{g^{(1)}})]^G+[\Delta
C_i^{j}(M_{g^{(1)}})]^\text{QED}+[\Delta C_i^{j}(M_{g^{(1)}})]^\text{EW},
\end{align}
with $i=1,2$ and $j=VLL, VRR, LR, RL$. Note that, in the $\text{RS}_c$ model, compared to the analogous
processes $K^0-\bar K^0$ and $B_s^0-\bar B_s^0$ mixings
\cite{Blanke:2008zb}, the $b\to ss{\bar d}$ decay receives
additional contributions from the $RL$ operators.
$[C_i^{j}(M_{g^{(1)}})]^G$ in Eq. (\ref{03}) denote the
contributions from the KK gluon to the Wilson coefficients that are
calculated to be
\begin{align}\label{04}
[C_1^{VLL}(M_{g^{(1)}})]^G=\frac{2}{3}{p_{\text{UV}}}^2\Delta_L^{sb}\Delta_L^{sd},\notag\\
[C_1^{VRR}(M_{g^{(1)}})]^G=\frac{2}{3}{p_{\text{UV}}}^2\Delta_R^{sb}\Delta_R^{sd},\notag\\
[C_1^{LR}(M_{g^{(1)}})]^G=-\frac{1}{3}{p_{\text{UV}}}^2\Delta_L^{sb}\Delta_R^{sd},\notag\\
[C_2^{LR}(M_{g^{(1)}})]^G=-2{p_{\text{UV}}}^2\Delta_L^{sb}\Delta_R^{sd},\notag\\
[C_1^{RL}(M_{g^{(1)}})]^G=-\frac{1}{3}{p_{\text{UV}}}^2\Delta_R^{sb}\Delta_L^{sd},\notag\\
[C_2^{RL}(M_{g^{(1)}})]^G=-2{p_{\text{UV}}}^2\Delta_R^{sb}\Delta_L^{sd},
\end{align}
where, $p_{\text{UV}}$ parameterizes the influence of brane kinetic
terms on the $\text{SU}(3)_c$ coupling. In our analysis we set
$p_{\text{UV}}\equiv1$. Similarly, for the KK photon and $(Z_H,
Z^{\prime})$ contributions, we find the following corrections to the
Wilson coefficients $C_i^{j}(M_{g^{(1)}})$,
\begin{align}\label{05}
[\Delta C_1^{VLL}(M_{g^{(1)}})]^\text{QED}&=2[\Delta_L^{sb}(A^{(1)})][\Delta_L^{sd}(A^{(1)})],\notag\\
[\Delta C_1^{VRR}(M_{g^{(1)}})]^\text{QED}&=2[\Delta_R^{sb}(A^{(1)})][\Delta_R^{sd}(A^{(1)})],\notag\\
[\Delta C_1^{LR}(M_{g^{(1)}})]^\text{QED}&=2[\Delta_L^{sb}(A^{(1)})][\Delta_R^{sd}(A^{(1)})],\notag\\
[\Delta C_2^{LR}(M_{g^{(1)}})]^\text{QED}&=0,\notag\\
[\Delta C_1^{RL}(M_{g^{(1)}})]^\text{QED}&=2[\Delta_R^{sb}(A^{(1)})][\Delta_L^{sd}(A^{(1)})],\notag\\
[\Delta C_2^{RL}(M_{g^{(1)}})]^\text{QED}&=0,
\end{align}
\begin{align}\label{06}
[\Delta C_1^{VLL}(M_{g^{(1)}})]^\text{EW}&=2[\Delta_L^{sb}(Z^{(1)})\Delta_L^{sd}(Z^{(1)})+\Delta_L^{sb}(Z_X^{(1)})\Delta_L^{sd}(Z_X^{(1)})],\notag\\
[\Delta C_1^{VRR}(M_{g^{(1)}})]^\text{EW}&=2[\Delta_R^{sb}(Z^{(1)})\Delta_R^{sd}(Z^{(1)})+\Delta_R^{sb}(Z_X^{(1)})\Delta_R^{sd}(Z_X^{(1)})],\notag\\
[\Delta C_1^{LR}(M_{g^{(1)}})]^\text{EW}&=2[\Delta_L^{sb}(Z^{(1)})\Delta_R^{sd}(Z^{(1)})+\Delta_L^{sb}(Z_X^{(1)})\Delta_R^{sd}(Z_X^{(1)})],\notag\\
[\Delta C_2^{LR}(M_{g^{(1)}})]^\text{EW}&=0,\notag\\
[\Delta C_1^{RL}(M_{g^{(1)}})]^\text{EW}&=2[\Delta_R^{sb}(Z^{(1)})\Delta_L^{sd}(Z^{(1)})+\Delta_R^{sb}(Z_X^{(1)})\Delta_L^{sd}(Z_X^{(1)})],\notag\\
[\Delta C_2^{RL}(M_{g^{(1)}})]^\text{EW}&=0,
\end{align}
where the overlap integrals $\Delta_{L,R}^{sb}(Z^{(1)})$,
$\Delta_{L,R}^{sb}(Z_X^{(1)})$, $\Delta_{L,R}^{sd}(Z^{(1)})$ and
$\Delta_{L,R}^{sd}(Z_X^{(1)})$ are given in Appendix B of
\cite{Blanke:2008zb}. These overlap integrals contain the profiles
of the zero mode fermions and shape functions of the KK gauge
bosons. We estimate the size of EW contributions
compared to the KK gluon contributions in the $b\to ss{\bar d}$
decay by factoring out all the couplings and charge factors from
$\Delta_{L,R}^{sb}$ and $\Delta_{L,R}^{sd}$. The remaining
$\tilde{\Delta}_{L,R}^{sb}$ and $\tilde{\Delta}_{L,R}^{sd}$ are then
universal for all the gauge bosons considered up to the different
boundary conditions. Combining contributions in Eqs. (\ref{04}),
(\ref{05}) and (\ref{06}) and evaluating the various couplings, we
have
\begin{align}\label{cont}
C_1^{VLL}(M_{g^{(1)}})=[0.67+0.02+0.56]\tilde{\Delta}_L^{sb}\tilde{\Delta}_L^{sd}=1.25\tilde{\Delta}_L^{sb}\tilde{\Delta}_L^{sd},\notag\\
C_1^{VRR}(M_{g^{(1)}})=[0.67+0.02+0.98]\tilde{\Delta}_R^{sb}\tilde{\Delta}_R^{sd}=1.67\tilde{\Delta}_R^{sb}\tilde{\Delta}_R^{sd},\notag\\
C_1^{LR}(M_{g^{(1)}})=[-0.333+0.02+0.56]\tilde{\Delta}_L^{sb}\tilde{\Delta}_R^{sd}=0.25\tilde{\Delta}_L^{sb}\tilde{\Delta}_R^{sd},\notag\\
C_1^{RL}(M_{g^{(1)}})=[-0.333+0.02+0.56]\tilde{\Delta}_R^{sb}\tilde{\Delta}_L^{sd}=0.25\tilde{\Delta}_R^{sb}\tilde{\Delta}_L^{sd},
\end{align}
where the three contributions in the bracket correspond to the KK
gluon, the KK photon and combined $(Z_H,Z^{\prime})$ exchange,
respectively. The Wilson coefficients $C_2^{LR}(M_{g^{(1)}})$ and
$C_2^{RL}(M_{g^{(1)}})$ receive only the KK-gluon contributions. We
see that the EW contributions, dominated by $(Z_H,Z^{\prime})$
exchanges, give +87$\%$ and +150$\%$ corrections in the case of
$C_1^{VLL}(M_{g^{(1)}})$ and $C_1^{VRR}(M_{g^{(1)}})$, respectively,
while corrections of $-$174$\%$ are observed for
$C_1^{LR}(M_{g^{(1)}})$ and $C_1^{RL}(M_{g^{(1)}})$. The Hamiltonian
in Eq. (\ref{01}) is valid at scales of $\mathcal{O}(M_{g^{(1)}})$
and has to be evolved to a low energy scale $\mu_b$ = 4.6 GeV. For
that, the anomalous dimension matrices for $\Delta F=2$ four-quark
dimension-six operators have already been calculated at two loop
level in \cite{Ciuchini:1997bw,Buras:2000if}. As gluons are flavor
blind and QCD preserves chirality so the anomalous dimension
matrices of the operators in $b\to ss{\bar d}$ are the same as for
the case of $B_{d,s}^0-\bar B_{d,s}^0$ mixing operators. Therefore,
the renormalization group running of the Wilson coefficients for the
$b\to ss{\bar d}$ decay is performed by using analytic formulae for
the relevant QCD factors given in Section 3.1 and appendix C of
\cite{Buras:2001ra}. Finally, the decay width for the $b\to ss{\bar
d}$ decay in the $\text{RS}_c$ model is given by
\begin{align}
\label{08}
{\it\Gamma}&=\frac{m_b^5}{3072(2\pi)^3(M_{g^{(1)}})^4}[16(|C_1^{VLL}(\mu_b)|^2+|C_1^{VRR}(\mu_b)|^2)\notag\\
&+12(|C_1^{LR}(\mu_b)|^2+|C_1^{RL}(\mu_b)|^2)+3(|C_2^{LR}(\mu_b)|^2+|C_2^{RL}(\mu_b)|^2)\notag\\
&-2\mathcal{R}e(C_1^{LR}(\mu_b)C_2^{*LR}(\mu_b)+C_2^{LR}(\mu_b)C_1^{*LR}(\mu_b)\notag\\
&+C_1^{RL}(\mu_b)C_2^{*RL}(\mu_b)+C_2^{RL}(\mu_b)C_1^{*RL}(\mu_b))].
\end{align}

\section{Bulk-Higgs RS model}\label{A3}
The bulk-Higgs RS model is based on the 5D gauge group
$\text{SU}(3)_c\times\text{SU}(2)_V\times\text{U}(1)_Y$, where all
the fields are allowed to propagate in the 5D space-time
\cite{Archer:2014jca}. $b\to ss{\bar d}$ decay in the bulk-Higgs RS
model results from tree-level exchanges of Kaluza-Klein gluons and
photons, the $Z^0$ boson and the Higgs boson as well as their KK
excitations and the extended scalar fields $\phi^{Z(n)}$. For the
bulk-Higgs RS model we consider the summation over the contributions
from the entire KK towers, with the lightest KK states having mass
$M_{g^{(1)}}\approx2.45$ $M_{\text {KK}}$. We start with the
effective NP Hamiltonian
\begin{align}
\label{09} [\mathcal{H}_{\text{eff}}^{\Delta
S=2}]_{\text{KK}}&=\sum_{n=1}^5[C_n\mathcal{O}_n+\tilde
{C}_n\tilde{\mathcal{O}}_n],
\end{align}
where
\begin{align}\label{10}
\mathcal{O}_1&=(\bar s_L\gamma_{\mu} b_L)(\bar s_L\gamma^{\mu}
d_L),\notag\\
\mathcal{O}_2&=(\bar s_R b_L)(\bar s_R d_L),\notag\\
\mathcal{O}_3&=(\bar s_R^{\alpha} b_L^{\beta})(\bar s_R^{\beta} d_L^{\alpha}),\notag\\
\mathcal{O}_4&=(\bar s_R b_L)(\bar s_L d_R),\notag\\
\mathcal{O}_5&=(\bar s_R^{\alpha} b_L^{\beta})(\bar s_L^{\beta}
d_R^{\alpha}).
\end{align}
A summation over color indices $\alpha$ and $\beta$ is understood. The $\tilde O_n$ operators are obtained from $O_n$ by
$L\leftrightarrow R$ exchange. Wilson coefficients at
$\mathcal{O}(M_{\text{KK}})$ are given by
\begin{align}\label{11}
C_1&=\frac{4\pi
L}{M_{\text{KK}}^2}(\tilde{\Delta}_D)_{23}\otimes(\tilde{\Delta}_D)_{21}[\frac{\alpha_s}{2}(1-\frac{1}{N_c})+\alpha Q_d^2+\frac{\alpha}{s_w^2 c_w^2}(T_3^d-Q_ds_w^2)^2],\notag\\
\tilde C_1&=\frac{4\pi
L}{M_{\text{KK}}^2}(\tilde{\Delta}_d)_{23}\otimes(\tilde{\Delta}_d)_{21}[\frac{\alpha_s}{2}(1-\frac{1}{N_c})+\alpha Q_d^2+\frac{\alpha}{s_w^2 c_w^2}(-Q_d s_w^2)^2],\notag\\
C_4&=-\frac{4\pi
L\alpha_s}{M_{\text{KK}}^2}(\tilde{\Delta}_D)_{23}\otimes(\tilde{\Delta}_d)_{21}-\frac{L}{\pi\beta M_{\text{KK}}^2}(\tilde{\Omega}_d)_{23}\otimes(\tilde{\Omega}_D)_{21},\notag\\
\tilde C_4&=-\frac{4\pi
L\alpha_s}{M_{\text{KK}}^2}(\tilde{\Delta}_d)_{23}\otimes(\tilde{\Delta}_D)_{21}-\frac{L}{\pi\beta M_{\text{KK}}^2}(\tilde{\Omega}_D)_{23}\otimes(\tilde{\Omega}_d)_{21},\notag\\
C_5&=\frac{4\pi
L}{M_{\text{KK}}^2}(\tilde{\Delta}_D)_{23}\otimes(\tilde{\Delta}_d)_{21}[\frac{\alpha_s}{N_c}-2\alpha Q_d^2+\frac{2\alpha}{s_w^2
c_w^2}(T_3^d-Q_ds_w^2)(Q_ds_w^2)],\notag\\
\tilde C_5&=\frac{4\pi
L}{M_{\text{KK}}^2}(\tilde{\Delta}_d)_{23}\otimes(\tilde{\Delta}_D)_{21}[\frac{\alpha_s}{N_c}-2\alpha Q_d^2+\frac{2\alpha}{s_w^2
c_w^2}(T_3^d-Q_ds_w^2)(Q_ds_w^2)],
\end{align}
where $Q_d=-1/3$, $T_3^d=-1/2$, and $N_c=3$.
Higgs and scalar field $\phi^Z$ give opposite contributions to the Wilson
coefficient $C_2$, thus they cancel each other giving $C_2=0$.
Similarly, $\tilde C_2=0$. The expressions of the mixing matrices
$(\tilde{\Delta}_{F(f)})_{mn}\otimes(\tilde{\Delta}_{F(f)})_{m^{\prime}n^{\prime}}$
and
$(\tilde{\Omega}_{F(f)})_{mn}\otimes(\tilde{\Omega}_{F(f)})_{m^{\prime}n^{\prime}}$
(with $F=U,D$ and $f=u,d,$ and similarly in the lepton sector) in terms
of the overlap integrals of boson and fermion profiles in the bulk-Higgs RS model, will be reported in \cite{Qin:2016jca}. For the
present study, we restrict ourselves to the $3\times3$ submatrices
governing the couplings of the SM fermion fields. In the zero mode
approximation (ZMA), the required expressions are simplified considerably
with (see also \cite{Bauer:2008xb})
\begin{align*}
(\tilde{\Delta}_{D})_{23}\otimes(\tilde{\Delta}_{d})_{21}&\rightarrow(U_d^{\dagger})_{2i}(U_d)_{i3}(\tilde{\Delta}_{Dd})_{ij}(W_d^{\dagger})_{2j}(W_d)_{j1},\notag\\
(\tilde{\Delta}_{Dd})_{ij}&=\frac{F^2(c_{Q_i})}{3+2c_{Q_i}}\frac{3+c_{Q_i}+c_{d_j}}{2(2+c_{Q_i}+c_{d_j})}\frac{F^2(c_{d_j})}{3+2c_{d_j}},\notag\\
(\tilde{\Omega}_{D})_{23}\otimes(\tilde{\Omega}_{d})_{21}&\rightarrow(U_d^{\dagger})_{2i}(W_d)_{j3}(\tilde{\Omega}_{Dd})_{ijkl}(W_d^{\dagger})_{2k}(U_d)_{l1},\notag\\
(\tilde{\Omega}_{Dd})_{ijkl}=&\frac{\pi(1+\beta)}{4L}\frac{F(c_{Q_i})F(c_{d_j})}{2+\beta+c_{Q_i}+c_{d_j}}\notag\\
&\times\frac{(Y_d)_{ij}(Y_d^{\dagger})_{kl}(4+2\beta+c_{Q_i}+c_{d_j}+c_{d_k}+c_{Q_l})}{4+c_{Q_i}+c_{d_j}+c_{d_k}+c_{Q_l}}\notag\\
&\times\frac{F(c_{d_k})F(c_{Q_l})}{2+\beta+c_{d_k}+c_{Q_l}},
\end{align*}
where $U_d$ and $W_d$ are flavor matrices diagonalising the SM
down-type Yukawa matrix. $\beta$ is a parameter of the model related
to the Higgs profile and $c's$ are bulk-mass parameters of fermions,
which control the localization of fermions in the warped extra
dimension. The 5D Yukawa matrix $Y_d$ has anarchic $\mathcal{O}(1)$
complex elements, which together with other flavor parameters
generate the right quark masses. Summation over indices $i, j, k$
and $l$ is understood. Analogous expressions hold for remaining
combinations of $D$ and $d$. The Effective Hamiltonian given in Eq.
(\ref{09}) is valid at $\mathcal{O}(M_{\text{KK}})$, which must be
evolved to a low-energy scale $\mu_b$. Hence for the evolution of
the Wilson coefficients we use the formulae of NLO QCD factors given
in \cite{Becirevic:2001jj}. After that, the decay width in the
bulk-Higgs RS model is given by
\begin{align}
\label{14}
{\it\Gamma}&=\frac{m_b^5}{3072(2\pi)^3}[64(|C_1(\mu_b)|^2+|\tilde C_1(\mu_b)|^2)\notag\\
&+12(|C_4(\mu_b)|^2+|\tilde C_4(\mu_b)|^2+|C_5(\mu_b)|^2+|\tilde C_5(\mu_b)|^2)\notag\\
&+4\mathcal{R}e(C_4(\mu_b)C_5^*(\mu_b)+C_4^*(\mu_b)C_5(\mu_b)\notag\\
&+\tilde C_4(\mu_b)\tilde C_5^*(\mu_b)+\tilde C_4^*(\mu_b)\tilde
C_5(\mu_b))].
\end{align}

\section{Phenomenological bounds on RS models}
In this section we discuss the
relevant constraints on the parameter spaces of the RS models coming from the EW precision tests
and the latest measurements of the Higgs signal strengths at the LHC. In addition, we will also consider
the constraints coming from $K^0-\bar K^0$ and $B_s^0-\bar
B_s^0$ mixing in Section \ref{sec:num}.

First, considering the $\text{RS}_c$ model, the bounds induced from
EW precision tests allow for KK masses in the few TeV range. A
recent tree-level analysis of the S and T parameters yields
$M_{g^{(1)}}>4.8$ TeV at $95\%$ confidence level (CL) for the mass
of the lightest KK gluon and photon resonances \cite{Malm:2013jia}.
While comparing the predictions of the signal rates for the various
Higgs-boson decays with the latest data from the LHC, it is
suggested in \cite{Malm:2014gha} that the most stringent bounds
emerge from the signal rates for $pp\rightarrow h\rightarrow
ZZ^{\ast}, WW^{\ast}$. In the RS$_c$ model, KK gluon masses lighter
than $22.7$ $\text{TeV}\times(y_{\star}/3)$ in the brane-Higgs case
and $13.2$ $\text{TeV}\times(y_{\star}/3)$ in the narrow bulk-Higgs
scenario are excluded at $95\%$ CL, where the
$y_{\star}=\mathcal{O}(1)$ is a free parameter and is defined as the
upper bound on the various entries of the Yukawa matrices that are
taken to be complex random numbers such that $|(Y_f)_{ij}|\le
y_{\star}$. Thus, for $y_{\star}=3$ the bounds derived from Higgs
physics are much stronger than those stemming from EW precision
measurements. In order to lower these bounds, smaller values of
$y_{\star}$ can be considered. For that it was also presented in
Ref. \cite{Malm:2014gha} that for the lowest value of the lightest
KK gluon mass $M_{g^{(1)}}=4.8$ TeV implied by EW precision
constraints, in the RS$_c$ model, the constraints at $95\%$ CL on
the values of the $y_{\star}$ are given by $y_{\star}<0.3$ for the
brane-Higgs scenario, and $y_{\star}<1.1$ for the narrow bulk-Higgs
case. However, realizing the fact that too small Yukawa couplings
would give rise to enhanced corrections to $\epsilon_K$ and hence
they would reinforce the RS flavor problem, relatively loose bound
on the values of the $y_{\star}$ can be obtained for the lightest KK
gluon mass of $M_{g^{(1)}}=10$ TeV. For instance, in the
$\text{RS}_c$ model, the constraints on the value of $y_{\star}$ at
$95\%$ CL valid for $M_{g^{(1)}}=10$ TeV are given by
$y_{\star}<1.1$ and $y_{\star}<2.25$ for the brane-Higgs case and
the narrow bulk-Higgs case, respectively \cite{Malm:2014gha}.

Next, we consider the bulk-Higgs RS model. The constraints on the KK
mass scale in the bulk-Higgs RS model implied by the analyses of EW
precision data are given in \cite{Archer:2014jca}. Under a
constrained fit (i.e. $U=0$), the obtained lower bounds on the KK
mass scale at $95\%$ CL vary between $M_{\text{KK}}>3.0$ TeV for
$\beta=0$ to $M_{\text{KK}}>5.1$ TeV for $\beta=10$. With an
unconstrained fit, these bounds relax to $M_{\text{KK}}>2.5$ TeV and
$M_{\text{KK}}>4.3$ TeV, respectively. For significantly larger
values of $\beta$, the lower bounds increase towards the brane
localized Higgs limit.

\begin{table}[ht]
\centering \caption{Default values of the input parameters used in the SM calculation \cite{Olive:2016xmw}.}
\begin{tabular}{|c|}
\hline
$G_{F}=1.16637\times 10^{-5}$ GeV$^{-2}$, $m_{b}=4.66_{-0.03}^{+0.04}$ GeV,\\
 $m_{c}=1.27\pm0.03$ GeV, $m_{t}=173.21\pm0.51$ GeV, $m_{W}=80.385\pm0.015$ GeV,\\
 $\tau_{B}=(1.566\pm0.003)\times 10^{-12}$ sec, $\sin{2\beta}=0.691\pm0.017$,\\
 $|V_{tb}V_{ts}^{\ast}|=(40\pm2)\times 10^{-3}$, $|V_{td}V_{ts}^{\ast}|=(32\pm3)\times
10^{-5}$, $|V_{cd}V_{cs}^{\ast}|=(21.8\pm0.6)\times
10^{-2}$.\\
\hline
\end{tabular}\label{table01}
\end{table}

\section{Numerical analysis}\label{sec:num}
In this section we present the results of the $b\to ss{\bar d}$ decay rate in RS models.
Before proceeding to analyze the NP, we first estimate the size of the leading
order SM result. The numerical values of the parameters that are involved in the SM
calculation are listed in table \ref{table01}. Employing the formula of the SM
$b\to ss{\bar d}$ decay rate \cite{Huitu:1998pa}, we get
\begin{align}
\label{15}
\mathcal{B}(b\to ss{\bar d})_{\text{SM}}=(2.19\pm0.38)\times10^{-12}.
\end{align}

\begin{figure}[th]
\center
\includegraphics[scale=0.5]{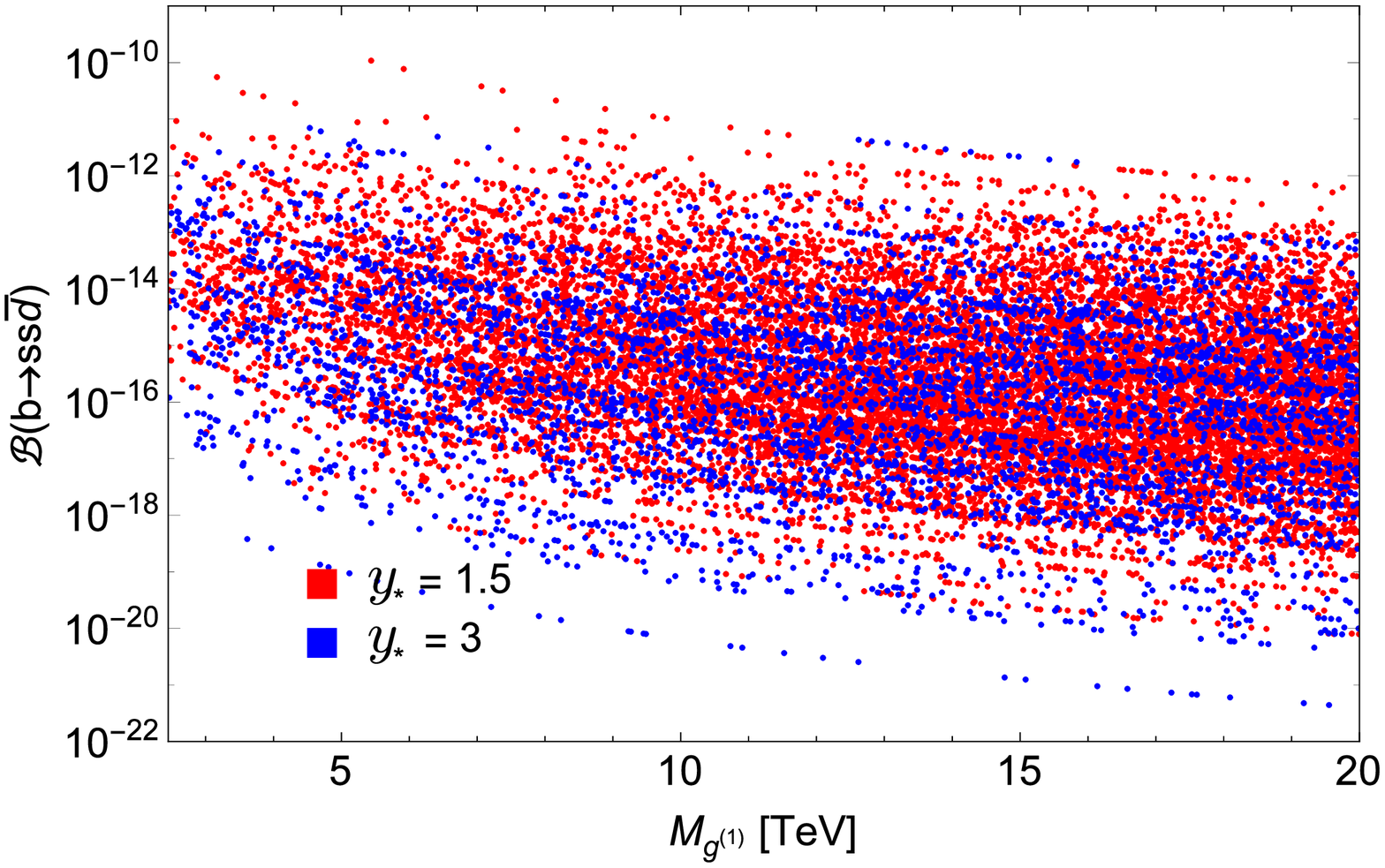}
\caption{The branching ratio of $b\to ss{\bar d}$ as a function of the KK gluon mass $M_{g^{(1)}}$ in $\text {RS}_c$ model. The red and blue points
correspond to $y_{\star}=1.5$ and $3$, respectively.}\label{fig:RSc}
\end{figure}

Next, we explore the parameter space of the $\text {RS}_c$ model by
the strategy outlined in \cite{Blanke:2008zb}. It was pointed out in
\cite{Blanke:2008zb} that there exist regions in parameter space,
without much fine-tuning in the 5D Yukawa couplings, which satisfy
all existing $\Delta F=2$ and EW precision constraints for scales of
masses of lightest KK gauge bosons $M_{\text{KK}}\simeq $ 3 TeV.
However, as mentioned above that for the anarchic Yukawa couplings
with $y_{\star}=3$ in the $\text {RS}_c$ model with the a brane
Higgs, the constraints on $M_{g^{(1)}}$ emerging from Higgs physics,
are much stronger than the EW precision constraints, so in our study
of the $\text {RS}_c$ model, we generate two sets of fundamental 5D
Yukawa matrices with $y_{\star}=1.5$ and $3$. For the first set the
28 parameters contained in the fundamental 5D Yukawa matrices are
randomly chosen in their respective ranges, $[0,\pi/2]$, $[0,2\pi]$
and $[0.1,1.5]$ for angles, phases and $|(Y_f)_{ij}|$, respectively.
Whereas, in the second set $|(Y_f)_{ij}|$ are chosen randomly in the
range $[0.1,3]$, by keeping ranges for angles and phases same as
previously. In order to determine the nine quark bulk-mass
parameters $c_{Q,u,d}^i$, we take $0.4\le c_Q^3\le 0.45$ in our
scan, allowing for consistency with EW precision data, so that the
remaining bulk mass parameters are determined making use of the
analytic formulae presented in section 3 of \cite{Blanke:2008zb}.
Finally, by diagonalising numerically the obtained effective 4D
Yukawa coupling matrices, we keep only those parameter sets that in
addition to the quark masses and CKM mixing angles also reproduce
the proper value of the Jarlskog determinant, all within their
respective $2\sigma$ ranges. The flavor transitions that would be
involved in the $b\to ss{\bar d}$ mode will commonly also give
contributions to $K^0-\bar K^0$ and $B_s^0-\bar B_s^0$ mixings, so
we consider $\Delta M_K$, $\epsilon_K$ and $\Delta M_{B_s}$
constraints on the parameter space in addition to EW precision
constraints and the Higgs constraints mentioned above. Expressions
of $(M_{12}^K)_{\text{KK}}$ and $(M_{12}^s)_{\text{KK}}$ relevant
for $K^0-\bar K^0$ and $B_s^0-\bar B_s^0$ mixings constraints,
calculated in the $\text {RS}_c$ model, are contained in Eqs. (4.32)
and (4.33) of \cite{Blanke:2008zb}, respectively. Figure
\ref{fig:RSc} shows the branching ratio of the RS$_c$ predictions
for the $b\to ss{\bar d}$ decay as a function of $M_{g^{(1)}}$ with
two different values of $y_{\star}$. Note that we have excluded the
SM contribution to display the decoupling behavior of the NP
contribution as $M_{g^{(1)}}$ increases. The red and blue scatter
points represent the cases of $y_{\star}=1.5$ and $3$, respectively.
While imposing the experimental constraints for $\Delta M_K$,
$\Delta M_{B_s}$ and $\epsilon_K$ in both cases, we set input
parameters in table \ref{table02} to their central values and allow
the resulting observables to deviate by $\pm 50\%$, $\pm 30\%$ and
$\pm 30\%$, respectively. The predictions of the $b\to ss{\bar d}$
decay rates for the parameter points with $y_{\star}=1.5$ are
generally larger than those with $y_{\star}=3$, but it can be seen
in figure \ref{fig:RSc} that after applying the $\Delta M_K$,
$\epsilon_K$ and $\Delta M_{B_s}$ constraints simultaneously, the
maximum possible $y_{\star}=1.5$ prediction is reduced relatively
close to that for the case of $y_{\star}=3$. However, after imposing
the $K^0-\bar K^0$ and $B_s^0-\bar B_s^0$ mixings constraints, still
for some parameter points with $y_{\star}=1.5$ in the low
$M_{g^{(1)}}$ range, the branching ratio of $b\to ss{\bar d}$ decay
in the $\text {RS}_c$ model can be close to the order of $10^{-10}$,
which is approximately two orders of magnitude larger compared to
the SM result. Considering the effects of the new heavy EW gauge
bosons $Z_H$ and $Z^{\prime}$ in the $\text {RS}_c$ model, we found
in agreement with \cite{Blanke:2008zb} that while imposing the
$\Delta M_K$ and $\epsilon_K$ constraints $Z_H$ and $Z^{\prime}$
give subleading contributions because the strong QCD renormalization
group enhancement of the $C_2^{LR}$ coefficient and the chiral
enhancement of the $\mathcal{Q}_2^{LR}$ hadronic matrix element in
$(M_{12}^K)_{\text{KK}}$ assure that the first KK gluon
contributions still dominate over EW contributions. However, for the
prediction of the branching ratio in the $b\to ss{\bar d}$ decay the
QCD renormalization group enhancement in the $C_2^{LR}$ and
$C_2^{RL}$ coefficients is smaller and the chiral enhancement is
absent. Therefore, for a parameter point that satisfies the $\Delta
M_K$, $\Delta M_{B_s}$ and $\epsilon_K$ constraints simultaneously,
$Z_H$ and $Z^{\prime}$ increase the prediction of the branching
ratio with comparable contributions to that of the first KK gluon.

\begin{table}[th]
\centering \caption{Values of experimental and theoretical
quantities used as input parameters while scanning the parameter
spaces of the RS models and in calculation of $\Delta M_K$, $\Delta
M_{B_s}$ and $\epsilon_K$. Values of the parameters $B_i^{K}$ at
$\mu_L=2$ GeV and $B_i^{s}$ at $\mu_b=4.6$ GeV are given in
$\overline{\text{MS}}$-NDR scheme obtained for $K^0-\bar K^0$ and
$B_s^0-\bar B_s^0$ mixings, respectively.}
\begin{tabular}{|l|l|l|}
\hline
$|V_{us}|=0.226(2)$& \multicolumn{2}{l|}{$s_w^2=0.2312$} \\
$|V_{ub}|=3.8(4)\times 10^{-3}$& \multicolumn{2}{l|}{$\alpha(m_Z)=1/127.9$}\\
$|V_{cb}|=4.1(1)\times 10^{-2}$ \hfill\cite{Bona:2006sa}& \multicolumn{2}{l|}{$\alpha_s(m_Z)=0.1182\pm0.0012$ \hfill\cite{Olive:2016xmw}} \\ \hline
$\lambda=0.2250\pm0.0005$ &\multicolumn{2}{l|}{ $m_K=497.611$ MeV}\\
$A=0.811\pm0.026$ &\multicolumn{2}{l|}{$m_{B_s}=5366.82$ MeV \hfill\cite{Olive:2016xmw}}\\ \cline{2-3}
${\bar\rho}=0.124_{-0.018}^{+0.019}$ & \multicolumn{2}{l|}{$\eta_{tt}=0.57\pm0.01$ \hfill\cite{Buras:1990fn}} \\
${\bar\eta}=0.356\pm0.011$ \hfill\cite{Olive:2016xmw}&\multicolumn{2}{l|}{$\eta_{cc}=1.50\pm0.37$ \hfill\cite{Herrlich:1993yv}} \\ \cline{1-1}
$\Delta M_K=(3.484\pm0.006)\times10^{-15}$ GeV &\multicolumn{2}{l|}{$\eta_{ct}=0.47\pm0.05$ \hfill\cite{Herrlich:1995hh,Herrlich:1996vf} }\\ \cline{2-3}
$\Delta M_{B_s}=(1.1688\pm0.0014)\times10^{-11}$ GeV &\multicolumn{2}{l|}{$\eta_B=0.55\pm0.01$ \hfill\cite{Buras:1990fn}} \\ \cline{2-3}
$|\epsilon_K|=(2.228\pm0.011)\times10^{-3}$  \hfill\cite{Olive:2016xmw}& \multicolumn{2}{l|}{$F_K=156$ MeV }\\ \cline{1-1}
$\phi_{\epsilon}=(43.52\pm0.05)^{\circ}$ \hfill\cite{Olive:2016xmw}& \multicolumn{2}{l|}{$F_{B_s}=245\pm25$ MeV  \hfill\cite{Lubicz:2008am}}\\
$\kappa_{\epsilon}=0.92\pm0.02$ \hfill\cite{Buras:2008nn}& \multicolumn{2}{l|}{}  \\ \hline
$\hat{B}_K=0.75$& $\mu_L=2$ GeV&$B_1^K=0.57,B_4^K=0.81,B_5^K=0.56$ \cite{Babich:2006bh} \\ \cline{2-3}
$\hat{B}_{B_s}=1.22$ \hfill\cite{Lubicz:2008am}&$\mu_b=4.6$ GeV &$B_1^s=0.87,B_4^s=1.16,B_5^s=1.75$ \hfill\cite{Becirevic:2001xt}\\
\hline
\end{tabular}\label{table02}
\end{table}

\begin{figure}[!ht]
  \centering
  \subfigure[$\beta$=1;]{
    \includegraphics[width=3.3in]{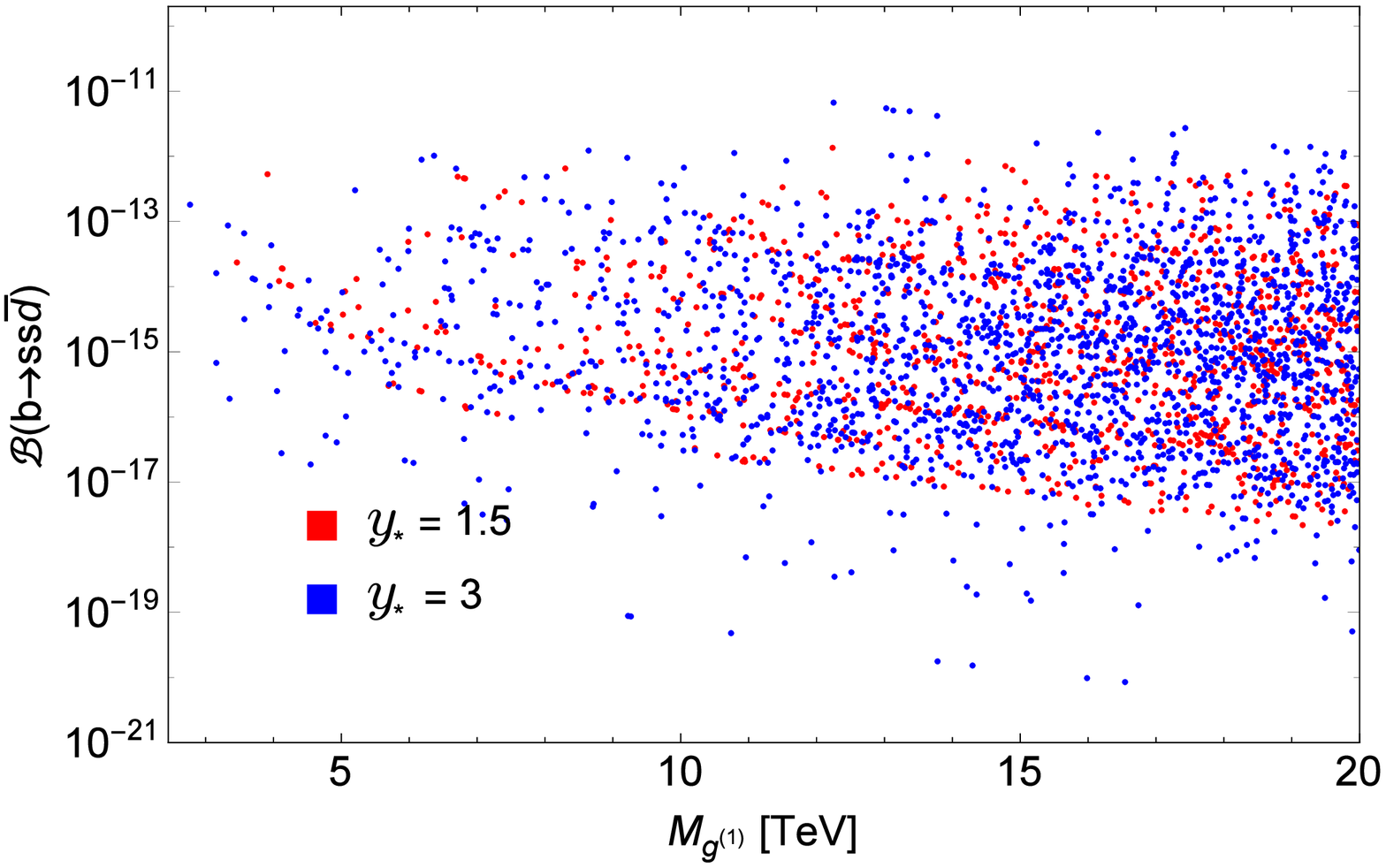}\label{fig:bHbeta1}}
  \hspace{0.in}
  \subfigure[$\beta$=10.]{
    \includegraphics[width=3.3in]{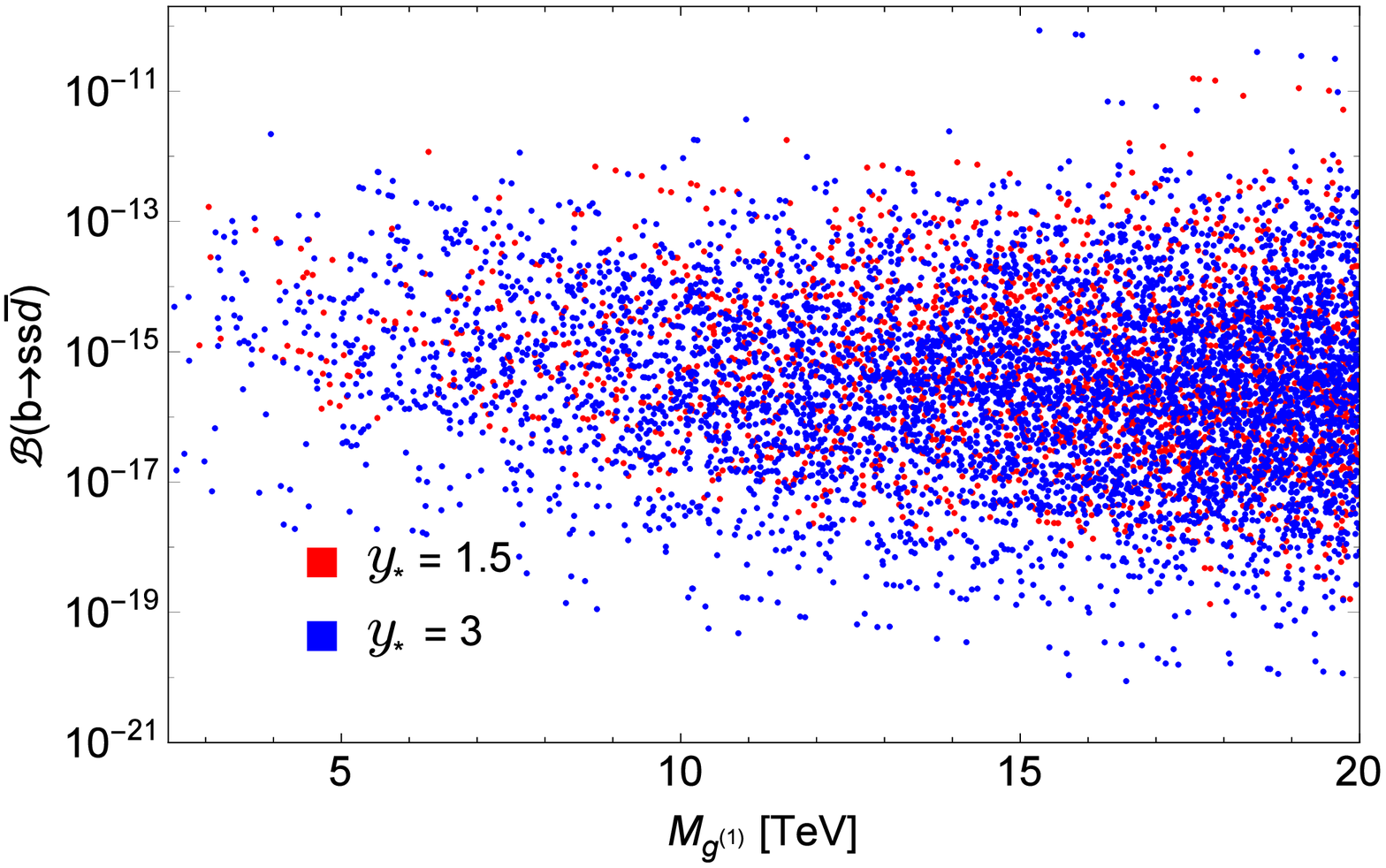}\label{fig:bHbeta10}}
  \caption{The branching ratio of $b\to ss{\bar d}$ as a function of the KK gluon mass $M_{g^{(1)}}$ in the bulk-Higgs RS model with $\beta=1$
  and $\beta = 10$. The red and blue scatter points correspond to $y_{\star}=1.5$ and $3$, respectively.}
  \label{fig:bH}
\end{figure}

For the bulk-Higgs RS model, following the directions given in
\cite{Archer:2014jca,Bauer:2009cf}, for a given value of $\beta$ and
$M_{\text {KK}}$, we generate two sets of random and anarchic 5D
Yukawa matrices, whose entries satisfy $|(Y_{u,d})_{ij}|\le
y_{\star}$ with $y_{\star}=1.5$ and 3. These values of $y_{\star}$
lie below the perturbativity bound, which is given by
$y_{\star}<y_{\text{max}}$ with
$y_{\text{max}}\sim8.3/\sqrt{1+\beta}$ \cite{Archer:2014jca}.
Moreover, for values of $y_{\star}<1$ it becomes increasingly
difficult to fit the top-quark mass. Next, we require that the 5D
Yukawa matrices with proper bulk-mass parameters $c_{Q_i}<1.5$ and
$c_{q_i}<1.5$ reproduce the correct values for the SM quark masses
evaluated at the scale $\mu=1$ TeV. In our analysis, we consider the
two representative values $\beta=1$ and $\beta=10$ corresponding to
broad Higgs profile and narrow Higgs profile, respectively. In
figure \ref{fig:bH}, we show the NP predictions with $\beta=1$ and
$10$, respectively, for the $b\to ss{\bar d}$ decay rate as a
function of $M_{g^{(1)}}$, after simultaneously imposing the $\Delta
M_K$, $\epsilon_K$ and $\Delta M_{B_s}$ constraints. The red and
blue scatter points again correspond to model points obtained using
$y_{\star}=1.5$ and 3, respectively. For the case of
$y_{\star}=1.5$, the branching ratios are generally larger because
of less suppressed FCNCs compared to $y_{\star}=3$ case, but as
mentioned earlier the lower values of $y_{\star}$ are subject to
more stringent constraints from flavour physics, so after imposing
the $\Delta M_K$, $\epsilon_K$ and $\Delta M_{B_s}$ constraints, the
maximum possible branching ratio of the parameter points with
$y_{\star}=1.5$ in the bulk-Higgs RS model lies close to the SM
result as shown in figure \ref{fig:bHbeta1}. While for the case of
$y_{\star}=3$ in figure \ref{fig:bHbeta1}, subject to relatively
less severe constraints from the $K^0-\bar K^0$ and $B_s^0-\bar
B_s^0$ mixings compared with $y_{\star}=1.5$ case, the maximum
possible branching ratio for some of the parameter points, even with
suppressed FCNCs, lies close to the order $10^{-11}$. Situation is
similar in the $\beta = 10$ case, except that compared to the
$\beta=1$ scenario, an order of magnitude enhancement for the
maximum possible branching ratio is observed for both cases of
$y_{\star}$, as displayed in figure \ref{fig:bHbeta10}.

\section{Conclusions}
\label{A4} We studied the $b\to ss{\bar d}$ decay in the
$\text{RS}_c$ and the bulk-Higgs RS model. In both models, main
contribution to the $b\to ss{\bar d}$ decay comes from tree level
exchanges of KK gluons, while in the $\text {RS}_c$ model the
contributions from the new heavy EW gauge bosons $Z_H$ and
$Z^{\prime}$ can compete with the KK-gluon contributions. We
employed renormalization group runnings of the Wilson coefficients
with NLO QCD factors in both models. Although this decay receives
tree level contributions, the parameter space is severely
constrained by $K^0-\bar K^0$ mixing and $B_s^0-\bar B_s^0$ mixing
experiments such that for broad Higgs profile corresponding to
$\beta=1$ case no significant increase in the branching ratio is
observed in the bulk-Higgs RS model compared to the SM result.
Whereas, for the value $\beta=10$, it is possible to achieve an
order of magnitude enhancement of the branching ratio for some of
the parameter points. While, the $\text {RS}_c$ model with
additional contributions from the new heavy EW gauge bosons $Z_H$
and $Z^{\prime}$ enhances the branching ratio, compared to SM
result, by at least one order of magnitude for some points in the
parameter space with $y_{\star}=1.5$, which leaves this decay free
for search of new physics in future experiments.

\section{Acknowledgements}
\label{A5} We are grateful to Wei Wang, Fu-Sheng Yu, Ying Li,
Si-Hong Zhou and Yan-Bing Wei for useful discussions. F.M. would
like to acknowledge financial support from CAS-TWAS president's
fellowship programme 2014. Q.Q. thanks the support from the UCAS-BHP
Billliton Scholarship. This work is supported in part by National
Natural Science Foundation of China under Grant No. 11375208,
11521505, 11235005 and 11621131001.
\bibliography{mybibfile}
\end{document}